\documentclass[aps,preprint]{revtex4}
\usepackage{graphicx}

\usepackage{bm}

\begin{document}

\title{ Anomalous Behavior of 2$^+$ Excitations around $^{132}$Sn
      }
\author{
J.~Terasaki,$^{1-3}$ J.~Engel,$^4$
W.~Nazarewicz,$^{1,2,5}$ and M.~Stoitsov$^{1-3,6}$ } 
\affiliation{
$^{1}$Department of Physics, University of Tennessee, 
Knoxville, Tennessee 37996, USA }
\affiliation{
$^{^2}$Physics Division, Oak Ridge National Laboratory, 
P.O.~Box 2008, Oak Ridge, Tennessee 37831, USA }
\affiliation{
$^{3}$Joint Institute for Heavy Ion Research, 
P.O.~Box 2008, Building 6008, MS 6374 
Oak Ridge, Tennessee 37831, USA }
\affiliation{
$^{4}$Department of Physics and Astronomy,
University of North Carolina, 
CB 3255, Phillips Hall, Chapel Hill, North Carolina 27599, USA }
\affiliation{
$^{5}$Institute of Theoretical Physics, University of Warsaw, 
ul.~Ho\.{z}a 69, PL-00-681 Warsaw, Poland }
\affiliation{
$^{6}$Institute of Nuclear Research and Nuclear Energy,
Bulgarian Academy of Science, Sofia-1784, Bulgaria
}
\date{ Sep.~27, 2002 }

\begin{abstract}

In certain neutron-rich Te isotopes, a decrease in the energy of the first
excited $2^+$ state is accompanied by a decrease in the $E2$ strength to that
state from the ground state, contradicting simple systematics and general intuition about
quadrupole collectivity.  We use a separable quadrupole{\-}-plus{\-}-pairing Hamiltonian
and the quasiparticle random phase approximation to calculate
energies, $B(E2,0^+\rightarrow 2^+)$ strengths, and $g$ factors for the lowest
$2^+$ states near $^{132}$Sn ($Z \geq 50$).  We trace the anomalous behavior in
the Te isotopes to a reduced neutron pairing  above the $N$ = 82 magic gap.
\end{abstract}
\maketitle

\section{Introduction}

As experiment pushes towards the nuclear drip line, it is
becoming possible to examine isotopic chains over increasingly large ranges
of $N$ and $Z$.  We have new opportunities to test systematics and
the ideas that underlie them.  One region in which experimental progress has been
made recently surrounds the neutron-rich doubly magic isotope $^{132}$Sn. 
In particular, recent Ref.~\cite{[Rad02]}
reports  measurements of 
the transition strengths
$B(E2;0^+\rightarrow 2^+)$ (or $B(E2)\!\!\uparrow$ for short)
from the ground state to the lowest 2$^+$ state for 
$^{132}$Te, 
$^{134}$Te, and  
$^{136}$Te. 
The authors discovered that $B(E2)\!\!\uparrow$'s and the energies of 
the lowest 2$^+$ states ($E_{2^+}$) behave differently in the
Te isotopes (with $N$ = 80, 82, and 84) than in those of Xe, Ba, and Ce which have
more protons.  In most isotopic chains, including those three, a decrease in
$E_{2^+}$
is accompanied by an increase in $B(E2)\!\!\uparrow$ as the states become
collective.  This is not the case in $^{132,136}$Te, where the
$B(E2)\!\!\uparrow$ decreases as $E_{2^+}$ decreases.

 Our work explains this
unusual behavior.  Our tool is the quasiparticle random phase approximation
(QRPA), in conjunction with a simple schematic interaction, which we apply to
even-even nuclei in the mass region $50\leq Z\leq 58$ and $80\leq N\leq 84$ (and
a much larger range of $N$ for the Sn chain).  The QRPA is a well-established
method for describing vibrational states \cite{[Kis63]} and has advantages of
simplicity, particularly when separable interactions are used and exchange terms
neglected. One should mention, that there exist large-scale shell-model calculations
for selected nuclei  around $^{132}$Sn \cite{[Cov02],[Rad02],[Ots02]}.  However,
at the present stage, these  calculations use different spaces (and
interactions) for nuclei above and below the $N$=82 magic gap. Our model,
albeit more phenomenological, uses the same Hamiltonian in both regions.

This paper is organized as follows:  In Sec.~II we review phenomenological
and simple microscopic approaches to the  systematics of
$E_{2^+}$ and $B(E2)\!\!\uparrow$.  In Sec.~III we give an overview of the
experimental data around $^{132}$Sn and discuss their significant properties.
In Sec.~IV we use the Hartree-Fock-Bogoliubov (HFB) method to discuss static
properties of the ground states.
The QRPA model is described in Sec.~V.
We show
results of the QRPA calculation for the lowest $2^+$ states in Sec.~VI and
discuss the origin of the irregular behavior of $B(E2)\!\!\uparrow$ from
a microscopic point of view in Sec.~VII.  The $g$ factors for Xe, Te, and Sn isotopes
are treated in Sec.~VIII.  Finally, Sec.~IX summarizes this work.

\section{Relation between $\bm{E_{2^+}}$ and $\bm{B(E2)\!\!\uparrow}$}

The systematic relation between $E_{2^+}$'s and $B(E2)\!\!\uparrow$'s is an old
topic.  One early phenomenological relation (by Grodzins \cite{[Gro62]}) is
\begin{equation}
 B(E2;0^+\rightarrow 2^+) = 14.9 \frac{1}{[E_{2^+}/{\rm keV}]}
 \frac{Z^2}{A} [e^2b^2] ,
\end{equation}
and another (by Raman {\em et al}.~\cite{[Ram01]}) is
\begin{equation}
 B(E2;0^+\rightarrow 2^+) = 3.26 \frac{1}{[E_{2^+}/{\rm keV}]}
 \frac{Z^2}{A^{0.69}} [e^2b^2] .
\end{equation}

The latter reproduces most of more than 300 experimental data points to within a factor
of
2.  Both these formulae, after factoring out a gentle dependence on $Z$ and $A$,
assert that $B(E2)\!\!\uparrow$'s are inversely proportional to $E_{2^+}$'s.
For vibrational states, this result is predicted, if mass parameter is 
constant, by the liquid drop model
\cite{[Boh75]}, which gives
\begin{equation}
B(E2;n_2=0\rightarrow n_2=1) =
5 \left( \frac{3}{4\pi}ZeR^2\right)^2
\frac{\hbar^2}{2D_2 E_{2^+}},
\end{equation}
where $R$ is the nuclear radius, and $D_2$
the quadrupole mass parameter. $n_2$ denotes the number of $2^+$ phonons.
It also falls
out of an RPA treatment of collective excitations in the simple microscopic
model of Brown and Bolsterli \cite{[Bro59]} and others \cite{[Hey90],[Rin80]}.
In physical terms,
collectivity lowers the energy of attractive modes while at the same time
increasing the transition strength because nucleons contribute coherently to the
transition.

\begin{figure}
\begin{center}
\includegraphics[width=0.5\textwidth]{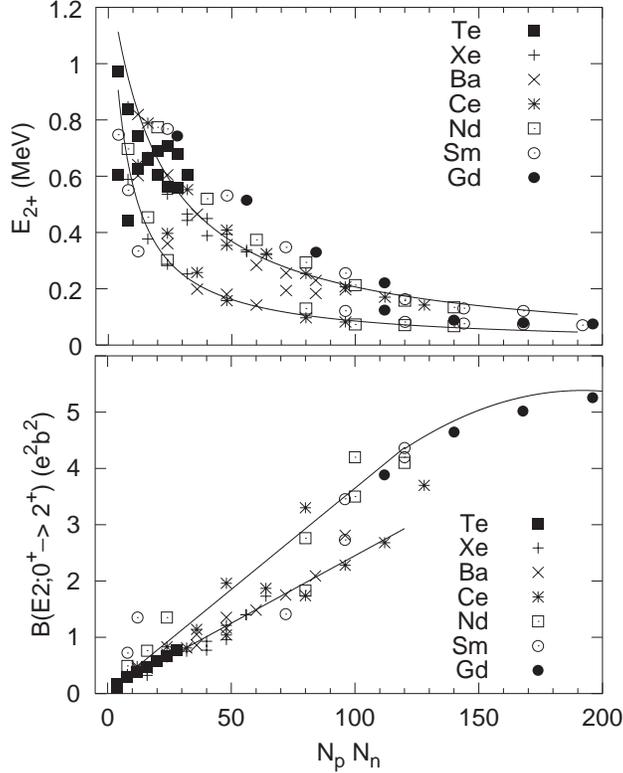}
\end{center}
\caption{Lowest $2^+$ energies (top) and $B(E2)\!\!\uparrow$'s 
(bottom) versus $N_pN_n$ in a number of even-even nuclei with $52 \leq Z\leq 64$.
The data are from Refs.\ \cite{[Rad02],[Ram01]}.
The curves are to guide the eye.}
\end{figure}

Another successful way of classifying collective $2^+$ states is the $N_pN_n$
scheme \cite{[Ham65],[Cas85],[Cas90a]}.  Both the $E_{2^+}$'s and
$B(E2)\!\!\uparrow$'s lie on smooth curves when plotted as functions of $N_p
N_n$, where $N_p$ $(N_n)$ is the number of valence proton (neutron)
particles or holes.  The plot for some nuclei
around
those considered in this work is shown in Fig.~1.  The data points can be divided,
approximately, into two well-correlated groups:  those for $N < 82$ (the upper $E_{2^+}$ and the
lower $B(E2)\!\!\uparrow$ branches) and those for $N > 82$ (the lower $E_{2^+}$
and the upper $B(E2)\!\!\uparrow$ branches).  The plots reveal a clear asymmetry in the
$2^+$ states with respect to $N = 82$.
That is, the $N > 82$ systems have {\it lower} $E_{2^+}$ and {\it higher}
$B(E2)\!\!\uparrow$ as compared to their $N < 82$ $N_pN_n$ partners.
This would suggest increased quadrupole collectivity in the region above $N > 82$.
However, as discussed in the following, 
deviation from this general trend can be found.

\section{Overview of data around $^{132}$Sn ($\bm{Z\geq 50}$)}
\begin{figure}
\begin{center}
\includegraphics[width=0.5\textwidth]{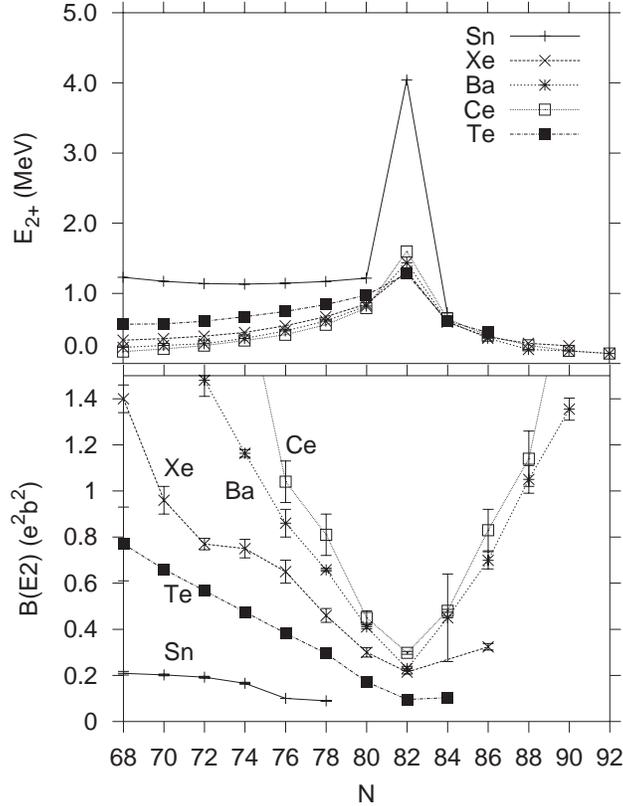}
\caption{Experimental values of
$E_{2^+}$ (top) and $B(E2)\!\!\uparrow$ (bottom) in
even-even Sn, Te, Xe, Ba, and Ce isotopes, 
as functions of neutron number $N$.
The experimental $B(E2)\!\!\uparrow$ rates were taken from 
Refs.~\cite{[Rad02],[Ram01],[Jak02]} (for $E_{2^+}$ cf.~\cite{[Ram01]}).  }
\end{center}
\end{figure}

Let us survey the experimental data relevant to this paper.
Figure 2 shows $E_{2^+}$'s and $B(E2)\!\!\uparrow$'s for the lowest $2^+$ states
of
even-even nuclei as functions of neutron number.
Both
observables are fairly symmetric around $N = 82$ for the Xe--Ce isotopes
indicating that particle and hole excitations in those nuclei play similar
roles.  Actually, some of the $B(E2)\!\!\uparrow$'s in Ce and Ba in the region
$N > 82$ are slightly larger than those with the same $N_n$ in $N < 82$;
similarly the $E_{2^+}$'s for $N > 82$ are lower than those for $N < 82$, in a
way consistent with the $N_pN_n$ plots of Fig.~1.  Clearly these isotopes
follow the usual relation between $B(E2)\!\!\uparrow$ and $E_{2^+}$.

On the other hand, $^{132}$Te, $^{134}$Te, and $^{136}$Te behave differently.
The $B(E2)\!\!\uparrow$ is not symmetric adjacent to $N$ = 82, a fact that is
even more significant when looking at the corresponding energies in Fig.~2. 
The state in $^{136}$Te lies 370 keV lower than that of $^{132}$Te, but nevertheless
the $B(E2)\!\!\uparrow$ in $^{136}$Te is smaller than that in $^{132}$Te.  The
situation violates the pattern of typical collective behavior discussed above.
[This behavior does not appear anomalous on the $N_pN_n$ plots of Fig.~1 because of the
scale of the figure,
however,  $N_pN_n$ = 4 for both $^{132}$Te and $^{136}$Te, and
$E_{2^+}(^{132}{\rm Te})$ = 0.974 MeV, $E_{2^+}(^{136}{\rm Te})$ = 0.606 MeV,
$B(E2,^{132}{\rm Te})$ = 0.172 $e^2b^2$, and $B(E2,^{136}{\rm Te})$ = 0.103
$e^2b^2$.]

%%%%%%%%%%%%%%%%%%%%%%%%%%%%%%%%%%%%%%% MARIO START %%%%%%%%%%%%%%%%%%%

\section{HFB calculations}
\begin{figure}
\includegraphics[width=0.8\textwidth]{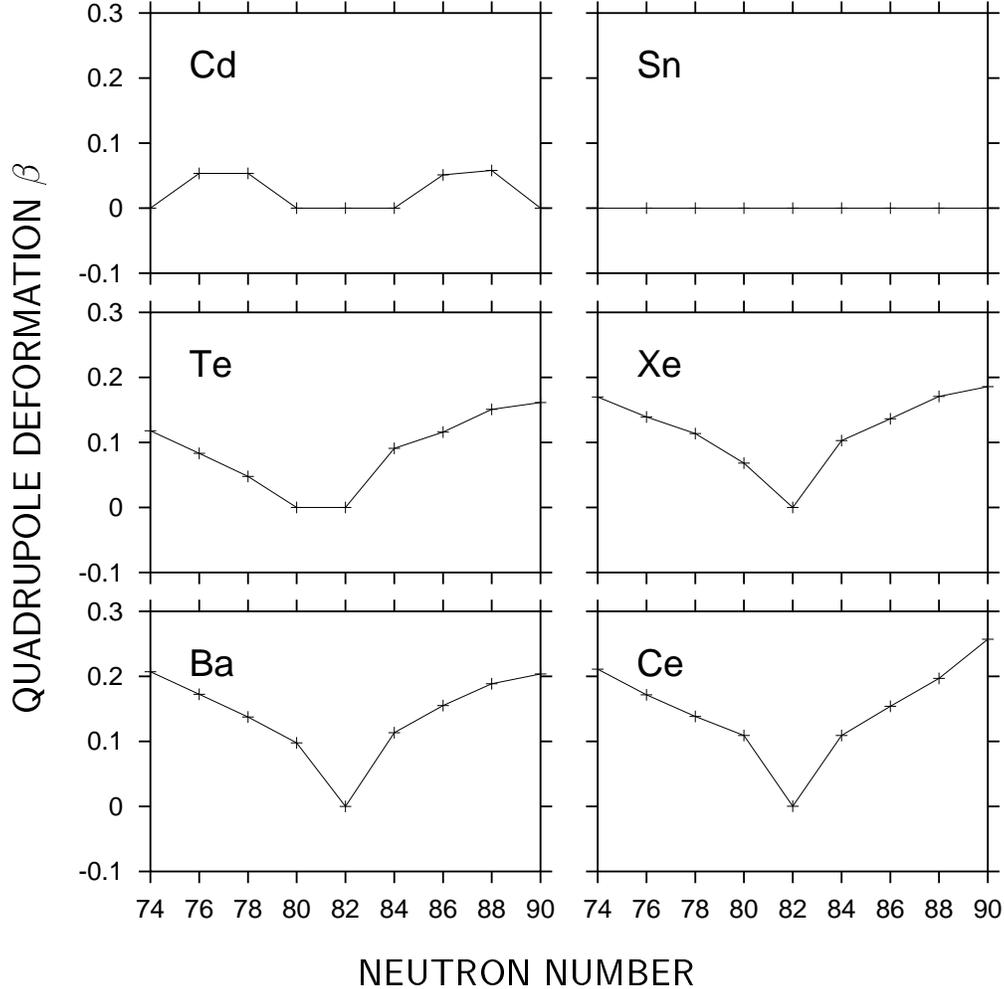}
\caption{The quadrupole deformation parameter $\beta$ calculated in the HFB
approximation with the Skyrme force SLy4 and an intermediate-type delta pairing
force \cite{[Dob02]}.}
\end{figure}

As a prelude to our QRPA treatment of the $2^+$ vibrations, we
calculate static shape and pairing deformations in the
Hartree-Fock-Bogoliubov (HFB) model of
Refs.~\cite{[Sto99],[Sto98b],[Sto98c],[Dob02],[Sto02]}.
We perform axially deformed HFB
calculations with the particle-hole Skyrme forces SLy4 \cite{[Cha97]} and an intermediate
contact delta pairing force \cite{[Dob02]}. The resulting quadrupole
deformation parameter
$\beta=\sqrt{\frac{\pi}{5}}\frac{1}{A}\frac{1}{R^2}Q$,
$Q$ being total quadrupole moment and R -- rms radius,
is shown in Fig.~3. It indicates
that the static deformation of the nuclei with $N$ = 80 and 84 is
zero or small ($\sim 0.1$) compared to those of the mid-shell nuclei.
We can therefore treat the $2^+$ states in these nuclei as
vibrations around a spherical shape.

In general, the HFB calculations follow the $N_pN_n$ trend
discussed earlier. The $\beta$ values above the $N = 82$ gap are
systematically increased for $N_n>4$. The strongest asymmetry
in the pattern of $\beta$ is predicted for the Te isotopes.

\begin{figure}
\begin{center}
\includegraphics[width=0.8\textwidth]{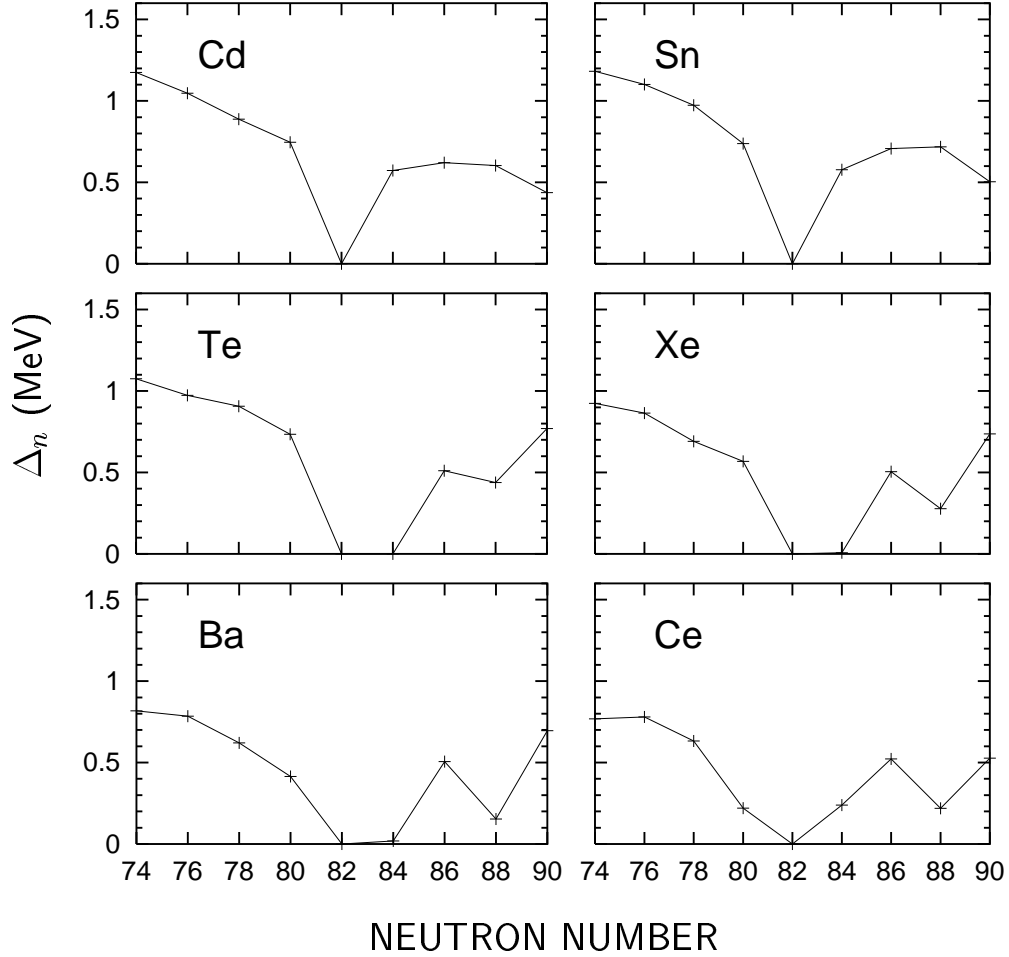}
\end{center}
\caption{Same as Fig.~3 but for the neutron pairing gaps.}
\end{figure}

Figure 4 shows predicted neutron pairing gaps.  
Since pairing is a symmetry-restoring interaction, 
the calculated pairing gaps are anticorrelated with the quadrupole
deformations.
Consequently, the values of $\Delta_n$ are systematically lower as
one crosses the $N=82$ gap. In particular, in most cases
$\Delta_n(N=80)>\Delta_n(N=84)$.

%%%%%%%%%%%%%%%%%%%%%%%%%%%%%%%%%%%%%%% MARIO END %%%%%%%%%%%%%%%%%%%

\section{QRPA calculation}

The Hamiltonian we use in our QRPA calculation is
\begin{equation}
H = \sum_\mu (\varepsilon_\mu - \lambda_{\tau}) c_\mu^\dagger c_\mu
-\sum_\tau \Delta_\tau (P_\tau^\dagger +P_\tau)
 + H_{\rm Q}^{\rm is}
+ H_{\rm Q}^{\rm iv}
 + H_{\rm Q}^{\rm p},
\label{Hamiltonian}
\end{equation}
where $\varepsilon_\mu$ is the single-particle energy, and $c_\mu^\dagger$ is 
the creation operator of a nucleon in the state $\mu$.
$\lambda_{\tau}$ is the chemical
potential, which depends on the isospin z-component $\tau$.
$\Delta_\tau$ is the pairing gap, and $P_\tau^\dagger$ is the
monopole pair creation operator.

As a residual two-body interaction, we use the sum of an isoscalar quadrupole
force $H_{\rm
Q}^{\rm is}$, an isovector quadrupole force $H_{\rm Q}^{\rm iv}$, and a quadrupole
pairing force $H_{\rm Q}^{\rm p}$, defined as follows:
\begin{eqnarray}
&& H_{\rm Q}^{\rm is} = -\frac{\chi_{T=0}}{2} \sum_m ({Q_m^{\rm pr}}^\dagger
+{Q_m^{\rm ne}}^\dagger ) (Q_m^{\rm pr} +Q_m^{\rm ne}),
\nonumber \\
&& H_{\rm Q}^{\rm iv} = -\frac{\chi_{T=1}}{2} \sum_m ({Q_m^{\rm pr}}^\dagger
-{Q_m^{\rm ne}}^\dagger ) (Q_m^{\rm pr}  -Q_m^{\rm ne} ),
\nonumber \\
&& Q_m^{\rm pr} = {\sum_{\mu\nu}}^{\rm proton} \langle\mu|r^2 Y_{2m}
|\nu\rangle c_\mu^\dagger c_\nu,
\nonumber \\
&& Q_m^{\rm ne} = {\sum_{\mu\nu}}^{\rm neutron} \langle\mu|r^2 Y_{2m}
|\nu\rangle c_\mu^\dagger c_\nu,
\nonumber \\
&& H_{\rm Q}^{\rm p} = -\sum_\tau\frac{G_2^\tau}{2}\sum_m {P_m^\tau}^\dagger
P_m^\tau,
\nonumber \\
&& {P_m^\tau}^\dagger = {\sum_{\mu\nu}}^\tau
 \langle\mu|r^2Y_{2m}|\nu\rangle c_\mu^\dagger c_{\bar{\nu}}^\dagger,
\nonumber\\
&& P_\tau^\dagger = {\sum_\mu}^\tau c_\mu^\dagger c_{\bar{\mu}}^\dagger,
\end{eqnarray}
where $\bar{\mu}$
denotes the time-reversed state of $\mu$.  For $\chi_{T=0}$, we use the
self-consistent values of Ref.~\cite{[Dob88]}; for $\chi_{T=1}$, we use the value
$\chi_{T=1} = \chi_{T=1}({\rm std}) =-92.9 A^{-7/3}$ MeV fm$^{-4}$.
(As will be seen later, the results of QRPA calculations are fairly insensitive
to the choice of $\chi_{T=1}$.)
We fix the quadrupole pairing strengths $G_2^\tau$ according to the prescription
proposed in
Ref.\ \cite{[Kub96]}. [We refer to this value as $G_2^\tau$(self).]
 Our QRPA equations are in the standard matrix form, as in
Ch.~14 of Ref.\ \cite{[Row70]}, and, as usual, we neglect the exchange terms of the
multipole-multipole interactions.

\begin{figure}
\begin{center}
\includegraphics[width=0.8\textwidth]{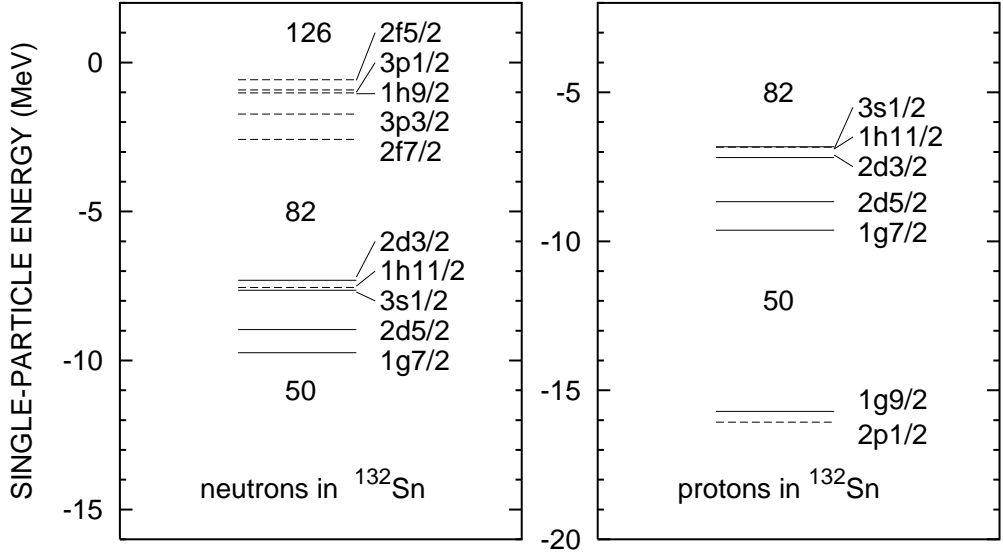}
\end{center}
\caption{The experimental s.p.~spectrum of $^{132}$Sn (from \cite{[Isa02]}).}
\end{figure}

Our calculations are performed in a single-particle (s.p.) space
of several harmonic-oscillator shells
($N_{\rm osc}$ = 2--6 for protons and $N_{\rm osc}$ = 2--7 for  neutrons).
Since
our configuration space is large, we use the bare, rather than effective, charges in
calculating $B(E2)\!\!\uparrow$. We take s.p.~energies $\varepsilon_\mu$ from experimental
data around $^{132}$Sn, shown in Fig.~5. 
(When the levels are not available this way, we
use Woods-Saxon energies \cite{[Cwi87]} for bound levels and Nilsson energies
\cite{[Nil69]} for unbound levels.) 
It is worth noting that the neutron level density just below the
82 shell gap is much larger than it is above the gap.
This is due to the near-degeneracy of $1h_{11/2}$, $2d_{3/2}$, and
$3s_{1/2}$ shells and a fairly large energy gap between the $2f_{7/2}$ and
$3p_{3/2}$ shells.
As we will see, this
difference plays a crucial role in the anomalous behavior of the Te isotopes.

\begin{figure}
\begin{center}
\includegraphics[width=0.8\textwidth]{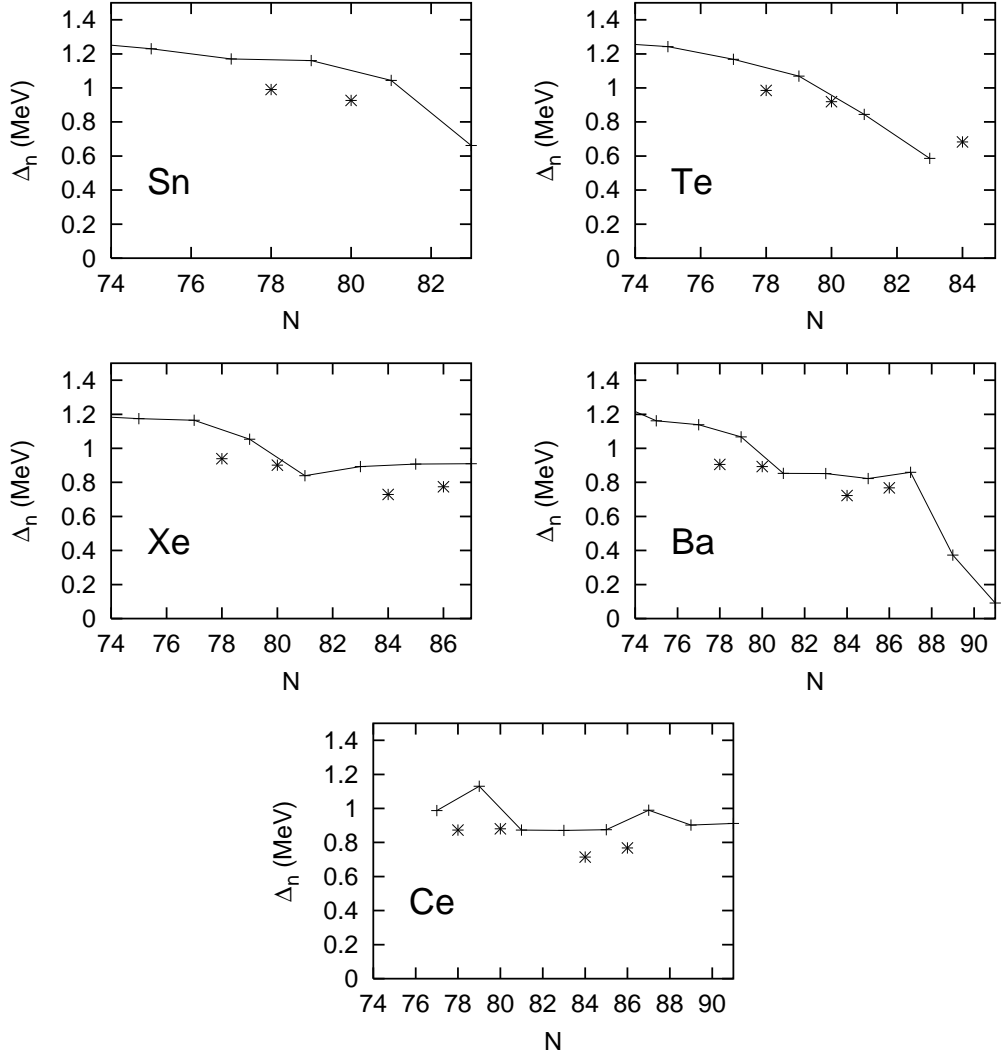}
\end{center}
\caption{The experimental neutron pairing gaps (connected by lines) obtained from
the odd-even mass differences and
calculated pairing gaps with the HFB-Lipkin-Nogami method (isolated symbols).
Experimental masses are from Ref.~\cite{[Aud95]}.}
\end{figure}

Figure 6 shows the experimental pairing gaps obtained from
odd-even mass differences, according to the prescription of Ref.~\cite{[Dob01]}, and
gaps calculated by the HFB-Lipkin-Nogami method \cite{[Dob02b]}.
We note that the HFB-Lipkin-Nogami calculation, which partly corrects for 
particle number fluctuations, reproduces
experimental trends very well.
The neutron
pairing gap in the Sn, Te, and Xe isotopes decreases as $N$ increases and
crosses $N$ = 82.
This effect, clearly seen also in the HFB calculation of Fig.~4, has been  noticed
earlier, cf.~Ref.~\cite{[Dob02]}.
In our QRPA calculations, we used renormalized experimental pairing gaps. 
The renormalization factors, reflecting the reduction of pairing in excited
$2^+$ states, were adjusted to experimental data in the Sn isotopes.
The renormalization factor turned out to be 0.6 (0.9) for neutrons (protons).
For magic nuclei with $N=82$ and/or $Z=50$, we took $\Delta$ =
0.4 MeV, a somewhat arbitrary value, reflecting the weak pairing correlations
in magic nuclei. (Experimental odd-even mass differences for magic nuclei do not determine
pairing gaps well \cite{[Dob01]}.) 
We used the average of the proton pairing gaps at $N=80$ and 84 for
$\Delta_p$ at $N=82$ to  avoid the sudden decrease at the magic number.

\section{Results of QRPA Calculations}
\begin{figure}
\includegraphics[height=0.9\textheight]{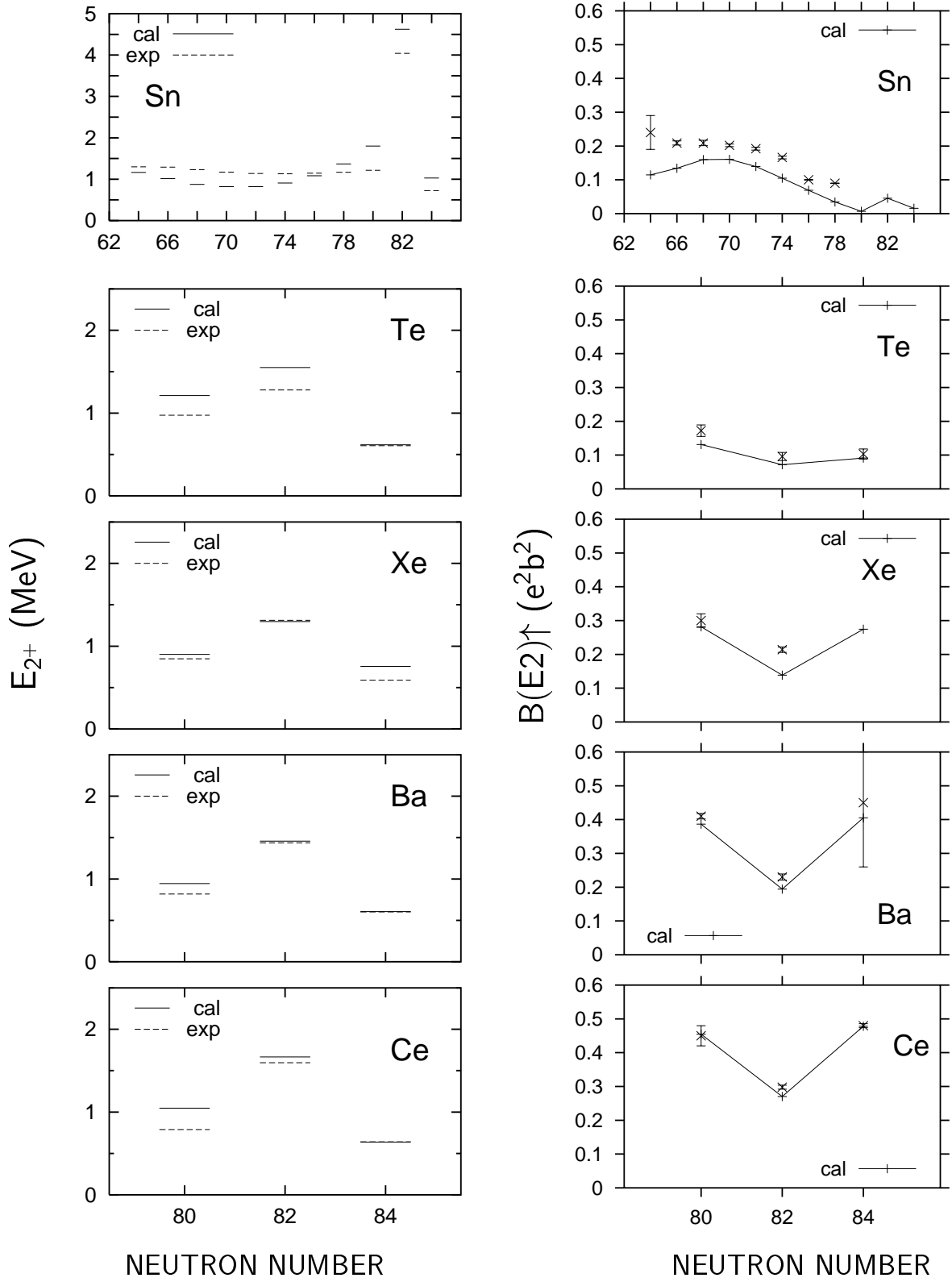}
\caption{$E_{2^+}$'s (left) and $B(E2)\!\!\uparrow$'s (right) from the QRPA
calculation and the experimental data.}
\label{e_be2_qrpa_nilsson}
\end{figure}
\mbox{}

\begin{figure}
\begin{center}
\includegraphics[width=0.8\textwidth]{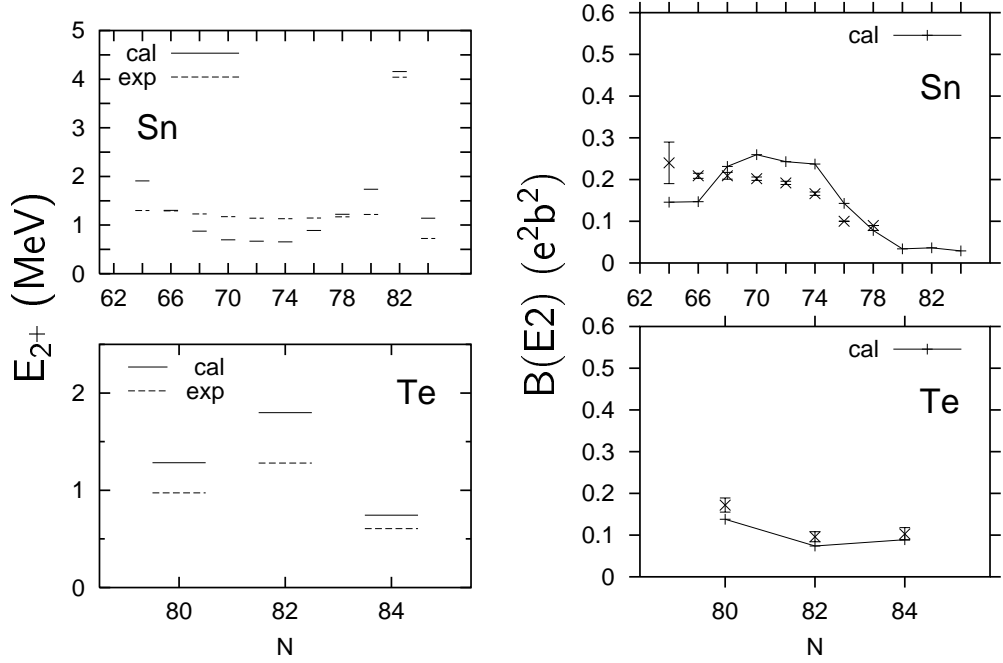}
\caption{Same as (part of) Fig.~7 but with the Nilsson single-particle energies.}
\end{center}
% \label{chi1-dep}
\end{figure}

\begin{figure}
\begin{center}
\includegraphics[height=0.4\textheight]{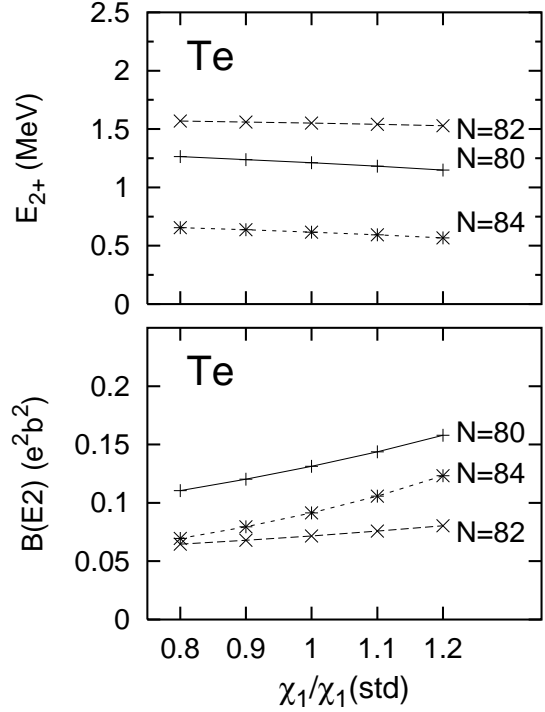}
\caption{Dependence of $E_{2^+}$ (top) and $B(E2)\!\!\uparrow$ (bottom) on
the strength of isovector quadrupole force $\chi_{T=1}$.
$\chi_{T=1}({\rm std}) = -92.9 A^{-7/3}$ MeV fm$^{-4}$.}
\end{center}
% \label{chi1-dep}
\end{figure}

\begin{figure}
\begin{center}
\includegraphics[height=0.4\textheight]{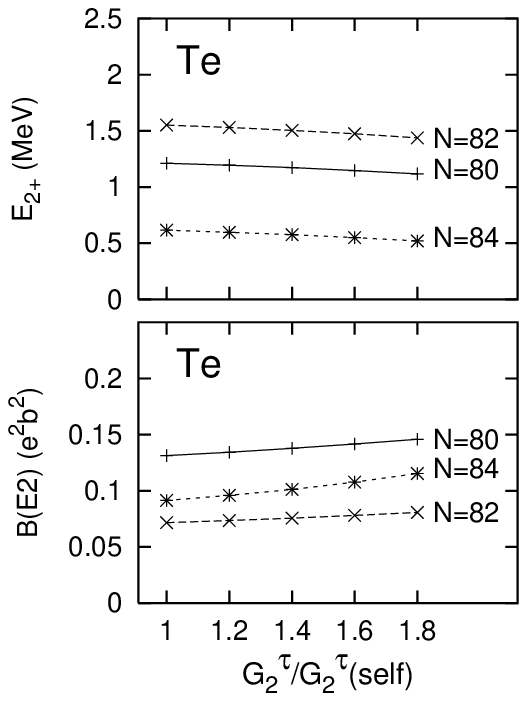}
\caption{Same as Fig.~9 but as functions of $G_2^\tau/G_2^\tau$(self).}
\end{center}
% \label{g2p-dep}
\end{figure}

We carried out QRPA calculations for even-$N$ isotopes of Sn with
$N$ = 64--84, and for the $N$ = 80, 82, 84 isotopes of Te, Xe, Ba, and Ce, which are
nearly spherical in our HFB calculations.
Figure 7 shows the calculated lowest 2$^+$ energies and
$B(E2)\!\!\uparrow$'s, along with the experimental data.  The calculations
reproduce the experimental trend quite well, in particular the asymmetry around
$N=82$ of the $B(E2)\!\!\uparrow$'s in the Te isotopes.  We also predict
an inverted, and more symmetric, curve for the $B(E2)\!\!\uparrow$'s in the Sn isotopes
with $N$ = 80--84.   This kind of inversion is well known to occur in the Pb region around
$N$ = 126 \cite{[Ram01]}.
(For more discussion of this point, see Sec.~VII.)
For comparison,
Fig.~8 shows the results with the pure Nilsson spectrum
(parameters from Ref.\ \cite{[Nil69]}). The collectivity in the $N$ = 68--76
isotopes of Sn is
enhanced here, but otherwise Figs.~7 and 8 are fairly similar.
Kubo {\it et al.}~\cite{[Kub96]} performed calculations in Sn isotopes up to $N$ = 74 with a
similar Hamiltonian and obtained similar results. In
the shell-model calculation of Ref.~\cite{[Rad02]}, $B(E2)\!\!\uparrow$ for
$^{134}$Te ($^{136}$Te ) turned out to be  0.088 (0.25) {\it $e^2 b^2$}, i.e.,
the transition rate has been predicted to increase when going from $N$=82
to $N$=84.

We checked the stability of our calculations by varying the strengths of the
isovector quadrupole force and the quadrupole pairing force.
Figures 9 and 10 show the results in Te.
The unusual behavior around $N=82$ clearly is not
sensitive to the strengths of these forces.
Based on all these results, we conclude that the QRPA prediction of the unusual
behavior around $^{136}$Te is robust and does not depend significantly on
model details, except for neutron pairing. 
%\begin{figure}
%\begin{center}
%\epsfig{file=Sn-e_x1_1.3.ps,height=4.0cm}
%\epsfig{file=Sn-be2_x1_1.3.ps,height=4.0cm}
%
%\epsfig{file=Te-e_x1_1.3.ps,height=4.0cm}
%\epsfig{file=Te-be2_x1_1.3.ps,height=4.0cm}
%\end{center}
%\caption{The 2$^+$ energies and $B(E2)\!\!\uparrow$ of Te isotopes
%with $\chi_{T=1}$ =  1.3 $\times $ the self-consistent value.}
%\end{figure}

\section{Abnormal pattern of Quadrupole Collectivity in the neutron-rich
Te isotopes}
\begin{figure}
\begin{center}
\includegraphics[height=0.6\textheight]{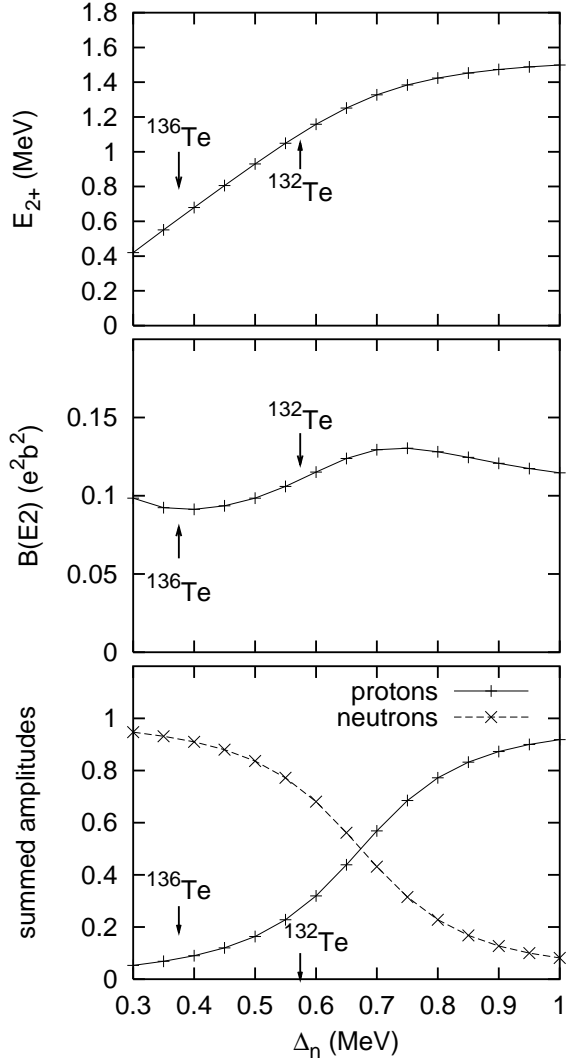}
\caption{The lowest $2^+$ energy (top), 
$B(E2)\!\!\uparrow$ (middle), and
the summed QRPA amplitudes
${\sum_{\mu\nu}}(\psi_{\mu\nu}^2-\varphi_{\mu\nu}^2)$ for protons and neutrons (bottom)
as functions of the neutron pairing gap in $^{136}$Te. The arrows show the
locations of the gaps in $^{132,136}$Te used in the solution in Fig.~7.}
\end{center}
\end{figure}

What is the reason for the unusual behavior of the Te isotopes around $N=82$,
i.e.\ the fact that {\it both} $E_{2^+}$ and the $B(E2)\!\!\uparrow$ are smaller in
$^{136}$Te than in $^{132}$Te?
The ingredient in our calculations that displays the most asymmetry around $N=82$ is the
neutron pairing gap.  To understand how it affects the results,
we performed QRPA calculations in $^{136}$Te for different values of
$\Delta_n$.
The results are shown in Fig.~11.
As $\Delta_n$ decreases from 0.6 MeV
to 0.4 MeV, both the $E_{2^+}$ and $B(E2)\!\!\uparrow$ decrease, indeed suggesting
that this quantity plays the key role in the unusual trend we want to explain.
To get more insight, we consider the forward ($\psi_{\mu\nu}$) and backward
($\varphi_{\mu\nu}$) QRPA amplitudes in the lowest-energy $2^+$ excitation
\begin{equation}
|2^+\rangle = \sum_{\mu < \nu}(\psi_{\mu\nu}a_\mu^\dagger a_\nu^\dagger
                 - \varphi_{\mu\nu}a_\nu a_\mu )|{\rm g.s.}\rangle,
\end{equation}
where $a_\mu^\dagger$ and $a_\mu$ create and annihilate a quasiparticle in the
state $\mu$, and $|{\rm g.s.\rangle}$ is the QRPA ground state.
The QRPA amplitudes $\psi_{\mu\nu}$ and $\varphi_{\mu\nu}$
depend on the ratios 
\begin{equation}
\frac{\langle\mu||Q^\tau||\nu\rangle}{{\cal E}_\mu + {\cal E}_\nu - E_{2^+}}
\:\:{\rm and}\:\:
\frac{\langle\mu||Q^\tau||\nu\rangle}{{\cal E}_\mu + {\cal E}_\nu + E_{2^+}} ,
\end{equation}
respectively, where ${\cal E}_\mu = \sqrt{(\varepsilon_\mu
- \lambda_\tau)^2 +\Delta_\tau^2}$
is the BCS quasiparticle energy.
The bottom portion of Fig.~11 shows
that these quantities depend significantly on the neutron pairing gap as well.

\begin{table}
\begin{center}
\begin{tabular}{rrrrcrrr}
\hline
\hline
\noalign{\vspace{1ex}}
 & $^{132}$Te & $^{134}$Te & $^{136}$Te &{~\strut} & $^{134}$Xe & $^{136}$Xe & $^{138}$Xe \\
\hline
\noalign{\vspace{1ex}}
$\displaystyle \sum_{\rm proton}\psi_{\mu\nu}^2$ & 0.63 & 0.99 & 0.12 &{~\strut}& 0.76 & 0.99 & 0.52\\
$\displaystyle \sum_{\rm neutron}\psi_{\mu\nu}^2$ & 0.44 & 0.02 & 0.97 &{~\strut}& 0.40 & 0.04 & 0.67\\
$\displaystyle \sum_{\rm proton}\varphi_{\mu\nu}^2$ & 0.03 & 0.00 & 0.04 &{~\strut}& 0.08 & 0.01 & 0.09\\
$\displaystyle \sum_{\rm neutron}\varphi_{\mu\nu}^2$ & 0.04 & 0.01 & 0.05 &{~\strut}& 0.08 & 0.02 & 0.10\\
\noalign{\vspace{1ex}}
\hline
\hline
\end{tabular}
\end{center}
\caption{Summed squared forward  ($\psi_{\mu\nu}^2$) and
backward  ($\varphi_{\mu\nu}^2$) QRPA amplitudes for $N$ = 80, 82, and 84 Te and Xe isotopes.}
\end{table}

The reason for the unusual behavior can be surmised from these figures.  The
decreased neutron pairing gap in $^{136}$Te means that the lowest neutron
quasiparticle energies are lower than  those in $^{132}$Te 
(0.792 MeV for $^{132}$Te and 0.460 MeV for $^{136}$Te).  As a result,
the energy of the lowest $2^+$ state
decreases when one crosses $N=82$.  But the low-lying neutron quasiparticle
energies also cause the neutron amplitudes in the wave function to increase and
the proton amplitudes to decrease, as Fig.~11 and Table I show.  Since
the $B(E2)\!\!\uparrow$ is determined solely by protons, it decreases as
well.
In other words, the
behavior of the lowest $2^+$ states reflects properties of the s.p.~spectrum
--- the fact that it is more dense below $N=82$ than above (see Sec.~V), giving
rise to a larger pairing gap --- more than collective quadrupole effects induced
by the residual interaction. This is not a total surprise given that both
isotopes have only 2 valence neutrons (or neutron holes).

\begin{figure}
\begin{center}
\includegraphics[height=0.6\textheight]{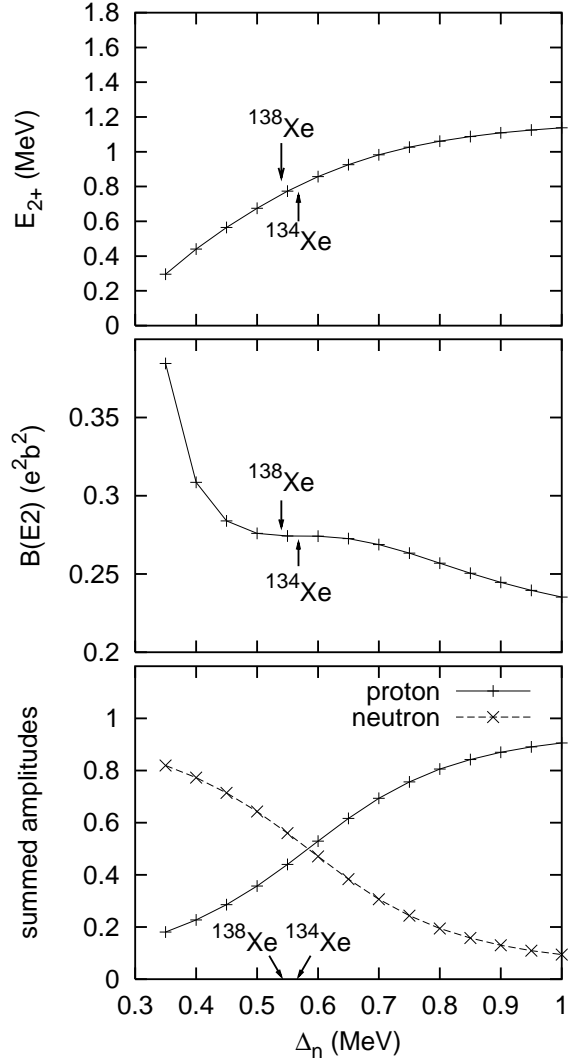}
\caption{Same as Fig.~11 but for 
$^{138}$Xe. The values of $\Delta_n$ in $^{134,138}$Xe, employed in QRPA calculations,
are marked by arrows.}
\end{center}
\end{figure}

 In the Xe, Ba, and
Ce, isotopes, the increased number of protons makes proton pairing and the
neutron-proton quadrupole-quadrupole interaction more important and reduces the
effectiveness of the s.p.~mechanism just described.
This is nicely illustrated in Fig.~12 for $^{138}$Xe.
One can see the usual relation between $E_{2^+}$ and $B(E2)\!\!\uparrow$ and 
a clear difference between Te and Xe in the $\Delta_n$-dependence of $B(E2)\!\!\uparrow$.
In Xe, $B(E2)\!\!\uparrow$ increases as the proton  amplitude decreases,  
indicating increased collectivity. 

The value of $B(E2)\!\!\uparrow$ in $^{134}$Te is smaller than that of $^{132}$Te,
in spite of 
the large proton amplitude (see Table I). 
However, the $2^+$ state in $^{134}$Te corresponds to one two-quasiparticle configuration $(g_{7/2})^2$, 
while the strength in  $^{132}$Te and $^{136}$Te is more fragmented, 
indicating the collective character of the $2^+$ state. 

\begin{figure}
\begin{center}
\includegraphics[width=0.5\textwidth]{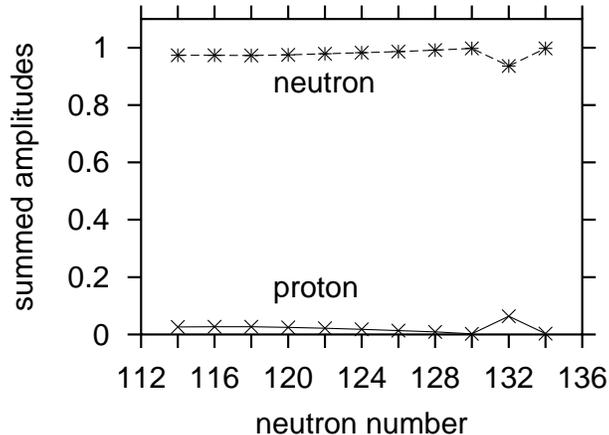}
\end{center}
\caption{Summed squared amplitudes
$\sum_{\mu\nu}(\psi_{\mu\nu}^2-\phi_{\mu\nu}^2)$ for 
the protons and neutrons of Sn isotopes.}
\end{figure}

We close this section by discussing the behavior of $B(E2)\!\!\uparrow$ 
of $^{130}$Sn--$^{134}$Sn mentioned in Sect.~VI (see Fig.~7). 
For this purpose, Fig.~13 shows summed QRPA amplitudes for protons and neutrons
in the Sn isotopes. 
It is clear that the neutron amplitudes are dominant in all cases. 
However, at $^{132}$Sn, {\it both} proton and neutron 
low-energy excitations are hindered; therefore the neutron amplitude 
decreases and the proton contribution increases, compared to the other isotopes.
This change causes a local increase in $B(E2)\!\!\uparrow$ at $^{132}$Sn.
(When the collectivity is small, $B(E2)\!\!\uparrow$ reflects the magnitude of
the proton amplitudes directly.)
Since the nucleus is in a neutron-rich region, however, 
matrix elements of the quadrupole operators of the neutrons 
are larger, on average, near the Fermi surface than those of the protons. 
Thus, excitations of the neutrons are still dominant in the $2^+$ state of 
$^{132}$Sn.

\section{$\bm{g}$ factors of Xe, Te, and Sn isotopes}

The abnormal behavior of the $E_{2^+}$'s and $B(E2)\!\!\uparrow$'s around $^{132}$Sn 
reflects the variations of proton and neutron amplitudes in the wave
function of the lowest $2^+$ state.
Therefore, we analyze the $g$ factor in neighboring nuclei; they are very
sensitive to relative proton/neutron compositions.

\begin{table}
\begin{center}
\begin{tabular}{llll}
\hline
\hline
\noalign{\vspace{1ex}}
& \hspace{1ex}$^{134}$Xe & \hspace{1ex}$^{136}$Xe & \hspace{0em}$^{138}$Xe \\
\noalign{\vspace{1ex}}
\hline
\noalign{\vspace{1ex}}
exp. & 0.354(7) & 0.766(45) & \\
cal. & 0.585    & 0.716     & 0.291  \\
% cal. & 0.427    & 0.539     & 0.305  \\
\noalign{\vspace{0.5ex}}
\hline
\hline
\end{tabular}
\caption{Experimental and calculated $g$ factors for $^{134,136,138}$Xe isotopes.
The data are from Ref.~\cite{[Jak02]}. }
\end{center}
% \end{table}

% \begin{table}
\begin{center}
\begin{tabular}{lllll}
\hline
\hline
\noalign{\vspace{1ex}}
& \multicolumn{2}{c}{neutron} & \multicolumn{2}{c}{proton}\\
& \mbox{\hspace{2.0ex}}fit & \hspace{1.5ex}th. &\hspace{2.0ex} fit &
\hspace{1.5ex}th.
\\
\hline
\noalign{\vspace{1ex}}
2$d_{3/2}$  & \hspace{1.8ex}0.554 & \hspace{1.8ex}0.534    & \hspace{1em}0.544 &
\hspace{1ex}0.419 \\
1$h_{11/2}$ &          $-$0.223   & $-$0.243 & \hspace{1em}1.39  &
\hspace{1ex}1.264 \\
3$s_{1/2}$  &          $-$2.65    & $-$2.674 & \hspace{1em}4.04  &
\hspace{1ex}3.906 \\
2$d_{5/2}$  &          $-$0.514   & $-$0.535 & \hspace{1em}1.54  &
\hspace{1ex}1.581 \\
1$g_{7/2}$  & \hspace{1.8ex}0.317 & \hspace{1.8ex}0.297 & \hspace{1em}0.803 &
\hspace{1ex}0.677  \\
\noalign{\vspace{1ex}}
\hline
\hline
\end{tabular}
\end{center}
\caption{The $g$ factors for neutron holes in $^{131}$Sn
and proton particles in $^{133}$Sb. 
The values labeled as ``fit" are taken from Ref.\ \cite{[Jak02]}, 
while the theoretical estimates are Schmidt values with $g_s$
multiplied by 0.7.}
\end{table}

We have calculated the $g$ factors of $^{134}$Xe,  $^{136}$Xe, and $^{138}$Xe,
and compare with recent data \cite{[Jak02]} in Table II.
As usual, we multiplied the bare spin $g_s$ factors by 0.7, and took bare 
$g_l$ factors \cite{[Boh75],[Rin80],[Hey90]}.
Our $g$ factor in $^{136}$Xe is larger than in $^{134}$Xe, though not by as much
as the data (see also Ref.\ \cite{[Kis63]}).
We show the corresponding proton and neutron QRPA amplitudes of 2$^+$ states in Table I.
Protons are more important in
$^{134}$Xe and $^{136}$Xe, while neutrons are more important in $^{138}$Xe.
We found by analyzing the amplitudes that the main component
of the 2$^+$ states of $^{134}$Xe and $^{136}$Xe is
$\pi$(1$g_{7/2}$)$^2$, while those of $^{138}$Xe are
$\pi$(1$g_{7/2}$)$^2$ and $\nu$(2$f_{7/2}$)$^2$.
It is interesting to compare the $g$ factors with those
of the single-particle states in Table III.
The observed $g$ factors for $^{134}$Xe and $^{136}$Xe support the idea that the
states of these nuclei consist mainly of
proton excitations (see Ref.\ \cite{[Jak02]});
our calculation is consistent with this picture.
The large $g$ factors of the proton 1$h_{11/2}$,
3$s_{1/2}$, and 2$d_{5/2}$ orbitals suggest that the nuclear $g$ factors
are sensitive to the small admixtures of these orbitals.  The Xe isotopes
therefore provide a severe test case of the many-body wave function.

\begin{table}
\begin{center}
\begin{tabular}{llll}
\hline
\hline
\noalign{\vspace{1ex}}
\hspace{1ex}$^{132}$Te & \hspace{1ex}$^{134}$Te & \hspace{0em}$^{136}$Te \\
\noalign{\vspace{1ex}}
\hline
\noalign{\vspace{1ex}}
 0.491 & 0.695 & $-$0.174 \\
\noalign{\vspace{0.5ex}}
\hline
\hline
\end{tabular}
\caption{The  calculated $g$ factors of $^{132,134,136}$Te isotopes.}
\end{center}
\end{table}

 Table IV displays calculated $g$ factors of the neutron-rich Te isotopes.
The  neutron dominance in
our $^{136}$Te wave function clearly lowers the predicted $g$ factor there. 
It would be interesting to test this prediction experimentally.

\begin{figure}
\begin{center}
\includegraphics[width=0.5\textwidth]{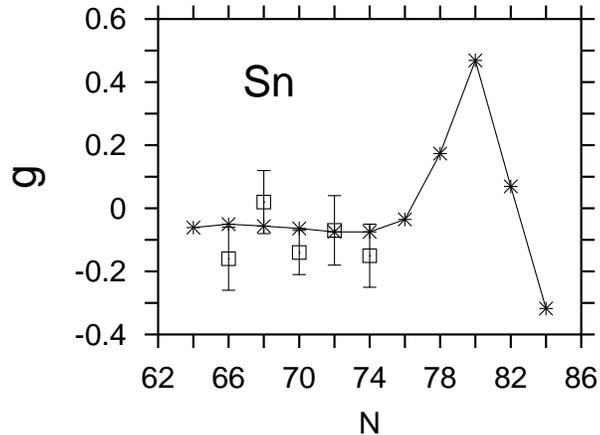}
\end{center}
\caption{Calculated (asterisks) and experimental~\cite{[Has80]} (open squares with error bars)
 $g$ factors of the lowest 2$^+$ states for Sn isotopes.
}
\end{figure}

 Finally, Fig.~14  shows calculated $g$ factors of Sn isotopes compared to
the experimental data. 
The behavior of the $g$ factors up to $N$ = 74 can be understood 
in terms of the negative single-neutron $g$-factors of the 1$h_{11/2}$,
2$d_{5/2}$, and 3$s_{1/2}$ shells (see Table~III and Ref.~\cite{[Has80]}).
 Around $N$=78, however, the 2$d_{3/2}$ orbital  carrying a 
 positive  $g$ factor becomes occupied, and this gives rise to positive
 $g$-factors in $^{128,130,132}$Sn. Above $N$=82, the structure of the lowest
 2$^+$ state is dominated by $2f_{7/2}$ shell, and $g$-factors drop again.

\section{Summary}
In this paper, we have investigated the irregular behavior of
$E_{2^+}$'s and $B(E2)\!\!\uparrow$'s in $^{132}$Te--$^{136}$Te through the QRPA with
a simple separable interaction.
%Prior to this, it turned out from the $N_pN_n$ plots that nuclei above $N = 82 $ have
%larger
%quadrupole collectivity in average than those below $N = 82$,
%and our HFB calculations show this tendency very well in $\beta$.
Our QRPA calculations reproduce the behavior seen in experiment,
and we trace the cause to the difference in neutron pairing below and above $N=82$.
The decrease in $\Delta_n$ with $N$ is clearly seen in experimental systematics and 
in self-consistent calculations. 
The results of our phenomenological model are fairly robust and depend only weakly
on other model parameters.
A related finding is that the
$B(E2)\!\!\uparrow$
in $^{132}$Sn should be larger than in its immediate Sn neighbors, as is the case around
$^{208}$Pb.
We hope that this prediction will stimulate further measurements in the
neutron-rich region around $^{132}$Sn.

To strengthen our argument about neutron dominance
in the wave function of the $2^+$ state in $^{136}$Te,
we also calculated $g$ factors of the Xe, Te, and Sn isotopes.
We reproduced the experimental trends and found
that while protons dominate the excitation amplitudes in
$^{134}$Xe and $^{136}$Xe,
the $g$ factor of the $2^+$ state of $^{136}$Te is
dramatically reduced. The experimental discovery of this effect as well as 
significant behavior of $^{128}$Sn--$^{134}$Sn   would
validate our understanding of the structure of nuclei around $^{132}$Sn.

\vspace{4em}

\begin{acknowledgments}
Discussions with C.~Baktash, D.C.~Radford, H.~Sakamoto, and K.~Matsuyanagi are gratefully acknowledged.
We are indebted to A.~Stuchbery for  information on the recently measured $g$ factors.
This work was supported in part by the U.S.~Department of Energy under Contract
Nos.~DE-FG02-96ER40963 (University of Tennessee), DE-AC05-00OR22725 with UT-Battelle,
LLC (Oak Ridge National Laboratory), and DE-FG02-97ER41019 (University of North
Carolina), and by the National Science Foundation Contract
 No.~0124053 (U.S.-Japan Cooperative Science Award).
\end{acknowledgments}

%\bibliography{Te-paper}
%\bibliographystyle{unsrt}

% \clearpage
\end{document}